# High-index dielectric metasurfaces performing mathematical operations

Andrea Cordaro[1,2], Hoyeong Kwon[3], Dimitrios Sounas[3,4], A. Femius Koenderink[2], Andrea Alù[3,5] and Albert Polman[2,*]

[1]Van der Waals-Zeeman Institute, Institute of Physics, University of Amsterdam
Science Park 904, 1098 XH Amsterdam, The Netherlands

[2]Center for Nanophotonics, AMOLF
Science Park 104, 1098 XG Amsterdam, The Netherlands

[3]Department of Electrical and Computer Engineering,
The University of Texas at Austin, Austin, TX 78712, USA.

[4]Department of Electrical and Computer Engineering,
Wayne State University, Detroit, MI 48202, USA

[5]Photonics Initiative, Advanced Science Research Center, City University of New York, New York, NY 10031, USA

[*]e-mail: a.polman@amolf.nl

**Image processing and edge detection are at the core of several newly emerging technologies, such as augmented reality, autonomous driving and more generally object recognition. Image processing is typically performed digitally using integrated electronic circuits and algorithms, implying fundamental size and speed limitations, as well as significant power needs. On the other hand, it can also be performed in a low-power analog fashion using Fourier optics, requiring however bulky optical components. Here, we introduce dielectric metasurfaces that perform optical image edge detection in the analog domain using a subwavelength geometry that can be readily integrated with detectors. The metasurface is composed of a suitably engineered array of nanobeams designed to perform either 1st- or 2nd-order spatial differentiation. We experimentally demonstrate the 2nd-derivative operation on an input image, showing the potential of all-optical edge detection using a silicon metasurface geometry working at a numerical aperture as large as 0.35.**

The amount of data that is being globally created, processed and stored is increasing at a remarkable pace. Furthermore, the advent of new technologies, such as augmented reality, autonomous driving, and many other emerging techniques, requires on-the-fly processing of large data files, such as images, at an increasing rate. Image processing is usually performed digitally but the speed and power consumption limits of standard microelectronic components have become a true bottleneck. Analog optical processing provides a promising route that may overcome these limitations.

The idea of computing optically can be traced back to the early 1960s prompted by pioneering work that exploited concepts of Fourier optics[1–3]. These elegant all-optical solutions, however, require bulky optical components that are not integrable into a larger nanophotonic system, and have remained unpractical. The unprecedented control of light propagation over a subwavelength thickness that has been recently enabled by optical metasurfaces opens entirely new opportunities for analog optical computing[4–9]. In fact, "computing metasurfaces" may benefit from the speed and low power consumption of optics while being amenable to on-chip integration, thus enabling hybrid optical and electronic data processing on a single chip.

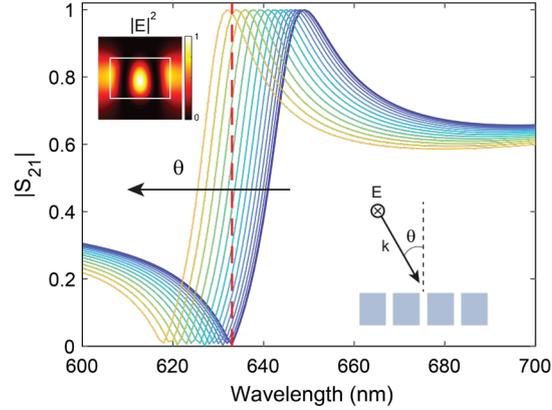

Fig. 1 | **Fano-resonant metasurface transmission spectra.** Simulated transmission spectra of a metasurface consisting of an array of dielectric nanobeams (width *w* = 182 nm, height *h* = 123 nm, pitch *p* = 250 nm and refractive index *n* = 4) as the incident angle is changed from 0 (blue line) to 0.3 rad (yellow line) in 15 steps. The red dashed line indicates the wavelength of operation ($\lambda$ = 633 nm). Insets: (bottom right) schematic of the proposed structure showing incoming light polarization; (top left) electric field amplitude profile within a unit cell at the resonant wavelength.

In this work, we design and realize optical metasurfaces composed of dielectric nanobeams that are illuminated by light polarized along the beams' direction. We tailor the spatial dispersion of the metasurfaces by controlling the leaky modes guided along the surface[10–12]. When the frequency and in-plane wave vector of incident light match one of these quasi-guided modes, an asymmetric Fano line-shape appears in the transmission spectrum[13–15] (see Figure 1), due to interference with the broad Fabry-Pérot resonance determined by the thickness and fill fraction of the structure. Figure 1 shows simulated transmission spectra of an array of dielectric nanobeams (width *w* = 182 nm, height *h* = 123 nm, and pitch *p* = 250 nm and refractive index *n* = 4 typical for Si) for incident angles ranging from 0 to 0.3 radians (17°). Due to the Fano interference, the transmission swings from 0 to unity within a narrow bandwidth. The sharp response in frequency corresponds to strong nonlocality: the spectrum is largely dependent on the incident angle and the transmission minimum shifts from $\lambda$ = 633 nm to $\lambda$ = 618 nm over the simulated angular range. The strong amplitude variation in transmission, and the sensitivity to the incoming k-vector, are often undesirable features of resonant metasurfaces, yet here these features enable the use of the metasurface as a Fourier spatial filter, and tailor with large flexibility its angular transmission response, or transfer function. In fact, by tuning the dispersion of the quasi-guided mode resonance, as well as the Fano line-shape asymmetry and linewidth, it is possible to design an optimized transfer function for a specific excitation wavelength (see Supporting Information) [15–21].

Starting with the case of 2$^{nd}$ order spatial differentiation, if f(x) is an arbitrary wave input signal, then its second derivative $\frac{d^2f(x)}{dx^2}$ equals $-k_x^2 \tilde{f}(k_x)$ in the spatial domain, where $\tilde{f}(k_x)$ is the Fourier transform of f(x). Therefore, 2nd order differentiation can be achieved by a metasurface that imparts a parabolic transfer function and a slowly varying phase to the spatial frequencies composing the input signal[22,23].

Figure 2a shows the simulated transmission amplitude $|S_{21}|$ and phase $\arg(S_{21})$ as a function of the in-plane wave vector $k_x$, normalized by the free space wavevector $k_0$ at the design wavelength $\lambda$ = 633 nm for the metasurface presented in Figure 1. In this design, optimized to perform 2nd-order differentiation, the angular response is close in amplitude to the ideal parabolic shape. The phase response shows a variation of approximately $0.9\pi$, deviating at high angles from the ideal constant phase response, but still providing a close-to-ideal second derivative response.

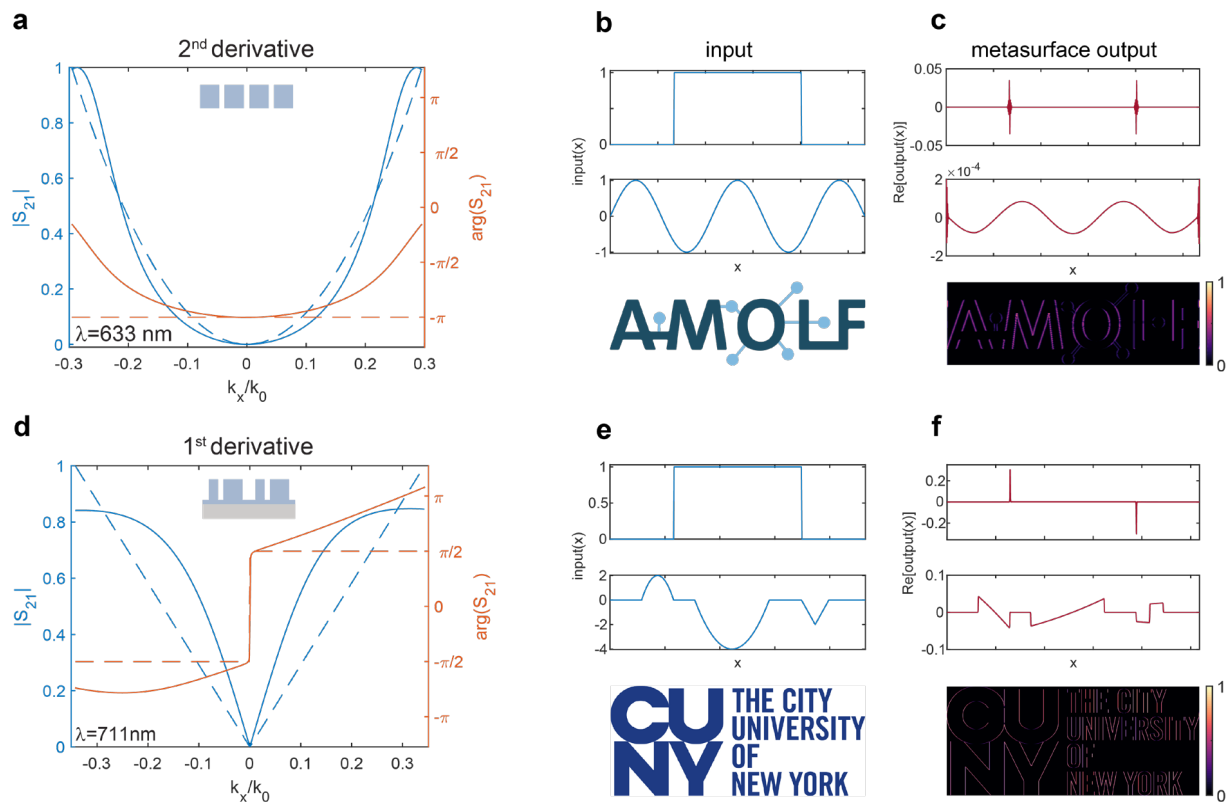

Fig. 2 | **Simulated transfer functions of dielectric metasurfaces performing 1st- and 2nd-order spatial differentiation. a** Transmission amplitude $|S_{21}|$ (solid blue line) and phase $\arg(S_{21})$ (solid orange line) for the metasurface optimized for 2nd-derivative operation (sketched in the inset) at $\lambda$ = 633 nm. The simulated transfer function is compared to the ideal case (dashed lines). The transmission reference plane is set such that the transmission phase at normal incidence equals $-\pi$. **b** Rectangular and sinusoidal input functions and 2D image that are used to numerically test the metasurface operation. The signal is discretized into 1000 pixels with individual pixel size set such that the Nyquist range matches the operational range in k-space of the metasurface. **c** Metasurface output for the input in b. For the 2D image, differentiation is performed line by line along the x-axis. **d-f** Same as a-c, but for 1st-derivative operation (metasurface geometry sketched in the inset) compared to the ideal case (dashed lines) at $\lambda$ = 711 nm. The transmission reference plane is set such that the transmission phase at normal incidence is 0.

Our metasurface design has two key features that distinguish it from earlier designs. First of all, the metasurface operational numerical aperture is large (*NA*≈0.35). This feature enables processing images with high spatial content and hence a resolution close to the diffraction limit. Moreover, it allows for direct implementation into standard imaging systems with similar NA, for instance by placing the metasurface right in front of a charge-coupled device (CCD) detector array, without needs for additional imaging lenses. This is a major advance over previously explored spatial differentiation schemes[6–8] that operate at an NA that is ~25 times smaller than what we demonstrate here. Second, the transmission in our design reaches unity at large wavevectors, enabling close-to-ideal image transformation efficiency, significantly larger than earlier attempts at realizing image processing metasurfaces.

Next, we use the optimized transfer function to numerically test how well the ideal 2$^{nd}$-order differentiation is approximated by our realistic metasurface design. Figure 2c shows the calculated response for rectangular and sinusoidal input functions shown in Figure 2b. The metasurface output clearly shows the edges of the rectangular input profile and flips the sinusoidal input function as expected. It is also possible to process arbitrary 2D images by performing the 2$^{nd}$ derivative line by line. The edges of one of our institutions' logos are clearly visible in Figure 2c. Notice that differentiation is performed only along the x-axis for this 1D geometry, hence the edges along the same direction are not detected.

To illustrate the flexibility of the metasurface image processing concept, Figure 2d shows the optimized transmission for a metasurface performing 1$^{st}$-order differentiation, which corresponds to the transfer function $S_{21}(k_x) = ik_x$ in the Fourier domain. In order to achieve such an operation with odd symmetry in space, we designed an asymmetric metasurface composed of an array of Si nanobeams with a unit cell (*p* = 300 nm) consisting of nanobeams with two different widths (*w*$_1$ = 48 nm, *w*$_2$ = 96 nm, *h* = 165nm and gap between the nanobeams 53 nm), placed on a thin silicon layer (thickness *t* = 35 nm) on a semi-infinite Al$_2$O$_3$ substrate. Indeed, by Lorentz reciprocity, it easy to prove that the unit cell has to be asymmetric both along the direction of propagation and transversally. Furthermore, the asymmetry in the phase response of the transfer function can be tuned with *t*, enabling the required π phase jump at k$_x$=0 (see Supporting Information)[5]. In this case, the experimental optical constants (including losses) for the two materials have been used in the simulations[24,25]. The simulated transfer function amplitude shows a linear behavior over a wavevector range up to k$_x$/k$_0$=0.1 (6°), above which it gradually bends away from the ideal case. For large angles, the transmission saturates below unity due to intrinsic absorption in Si. Figure 2e-f shows the calculated metasurface output for rectangular, parabolic and triangular input functions. The input slope changes and a nearly linear derivative for the parabola are clearly resolved. Furthermore, processing the logo results in clear detection of the edges in the x-direction consistently with 1$^{st}$-derivative operation.

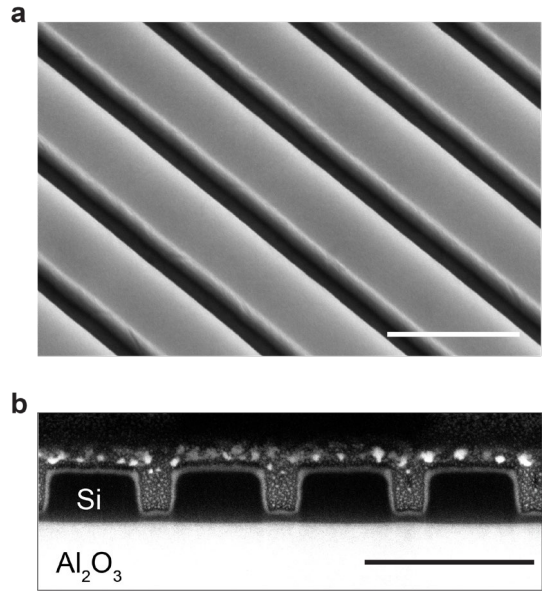

Fig. 3 | **Experimental Si metasurface performing 2$^{nd}$-order spatial differentiation. a** Tilted SEM image of the Si metasurface performing the 2$^{nd}$-derivative operation. **b** SEM image of a FIB cross section of the same metasurface showing the Si nanobeams on an Al$_2$O$_3$ substrate. The scale bar is 400 nm for both panels.

In order to demonstrate 2$^{nd}$-order differentiation experimentally, we fabricated a 1D array of Si nanobeams on a 0.46 mm thick sapphire (Al$_2$O$_3$) substrate, using a combination of electron beam lithography and reactive ion etching (see Methods). Figure 3a depicts a scanning electron microscopy (SEM) image of the fabricated structure showing high uniformity over a large area. The optimized dimensions for this design are $w$ = 206 nm, $h$ = 142 nm, and $p$ = 300 nm. Figure 3b shows a cross-section of the nanobeams. The etched Si sidewalls are straight to within ∼20 nm and a thin residual Si layer (thickness $t$ ≈ 22nm) is intentionally left between the pillars. This layer is essential to achieve optimum transmission for large wavevectors.

Figure 4a shows the measured transmittance (T = $|S_{21}|^2$) spectra as the incident angle is changed from 0 to 25°. In agreement with the simulated data in Figure 1, the Fano resonance shifts to shorter wavelengths as the angle is increased. The transmittance minimum is observed at $\lambda$ = 726 nm for normal incidence and amounts to 2.2%, the residue attributed to minor fabrication imperfections. Figure 4b shows the transmittance as a function of the in-plane wavevector at $\lambda$ = 726 nm, derived from the data in Figure 4a. The corresponding transmission amplitude ($|S_{21}|$) derived from the data is also plotted, along with the ideal parabolic amplitude response. The overall trend with increasing transmittance as a function of angle is well reproduced experimentally, with a significant residual transmittance at normal incidence and a maximum amplitude at largest angle of 0.84, which is mostly determined by the absorption in Si. Employing alternative high-index materials could further enhance the transmission for large angles.

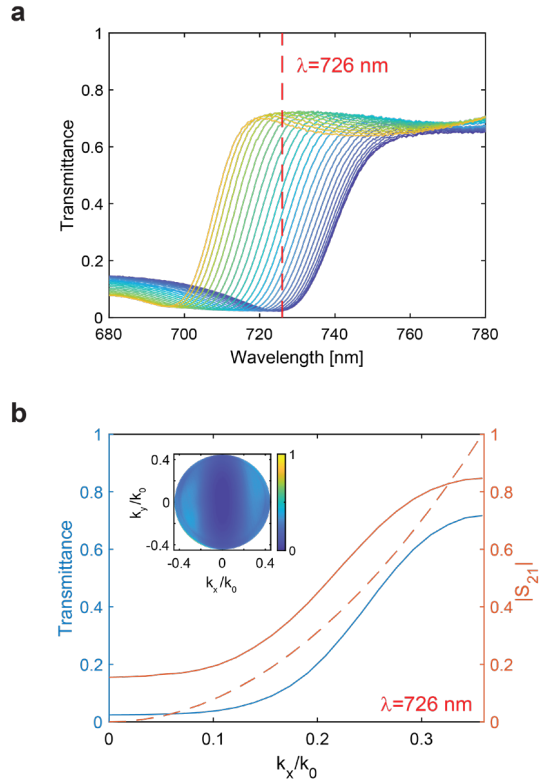

Fig. 4 | **Experimental metasurface transmission. a** Measured transmission spectra of the metasurface in Fig. 3 as the angle of incidence is increased from 0 (blue line) to 25° (yellow line) in 25 steps. **b** Measured transmittance (blue line) and corresponding calculated (Transmittance = $|S_{21}|^2$) transmission amplitude $|S_{21}|$ (orange solid line) as function of incident in-plane wave vector $k_x/k_0$ at $\lambda$ = 726 nm. The dashed orange line shows the ideal parabolic transfer function for $|S_{21}|$. Inset: measured metasurface transfer function amplitude at $\lambda$ = 726 nm (see Supporting Info for the description of the setup used).

The inset in Figure 4b shows the metasurface transfer function amplitude in wave number space taken with a Fourier microscope (see Supporting Information). These measurements clearly show the 1D nature of the metasurface operation. In fact, low $k_x$ spatial components are suppressed also for a wide range of $k_y$. Hence, the 2nd-order differentiation is experimentally performed in a line-by-line fashion, in agreement with the numerical calculation shown in Figure 2.

Finally, we experimentally investigate the 2nd-derivative operation of the Si metasurface when an image is projected onto the sample. To this end, we fabricated one of our institutions' logos using a Cr pattern on glass (Figure 5a). We first project the image onto the metasurface using off-resonant illumination ($\lambda$ = 750 nm) and then image the metasurface output onto a CCD imaging camera (Figure 5b) (see Supporting Information); the contrast of the input object is clearly maintained. On the other hand, for resonant illumination at $\lambda$ = 726 nm (Figure 5c) the edges are clearly resolved in the transformed image. As expected, no edge contrast is observed for features along the x-direction.

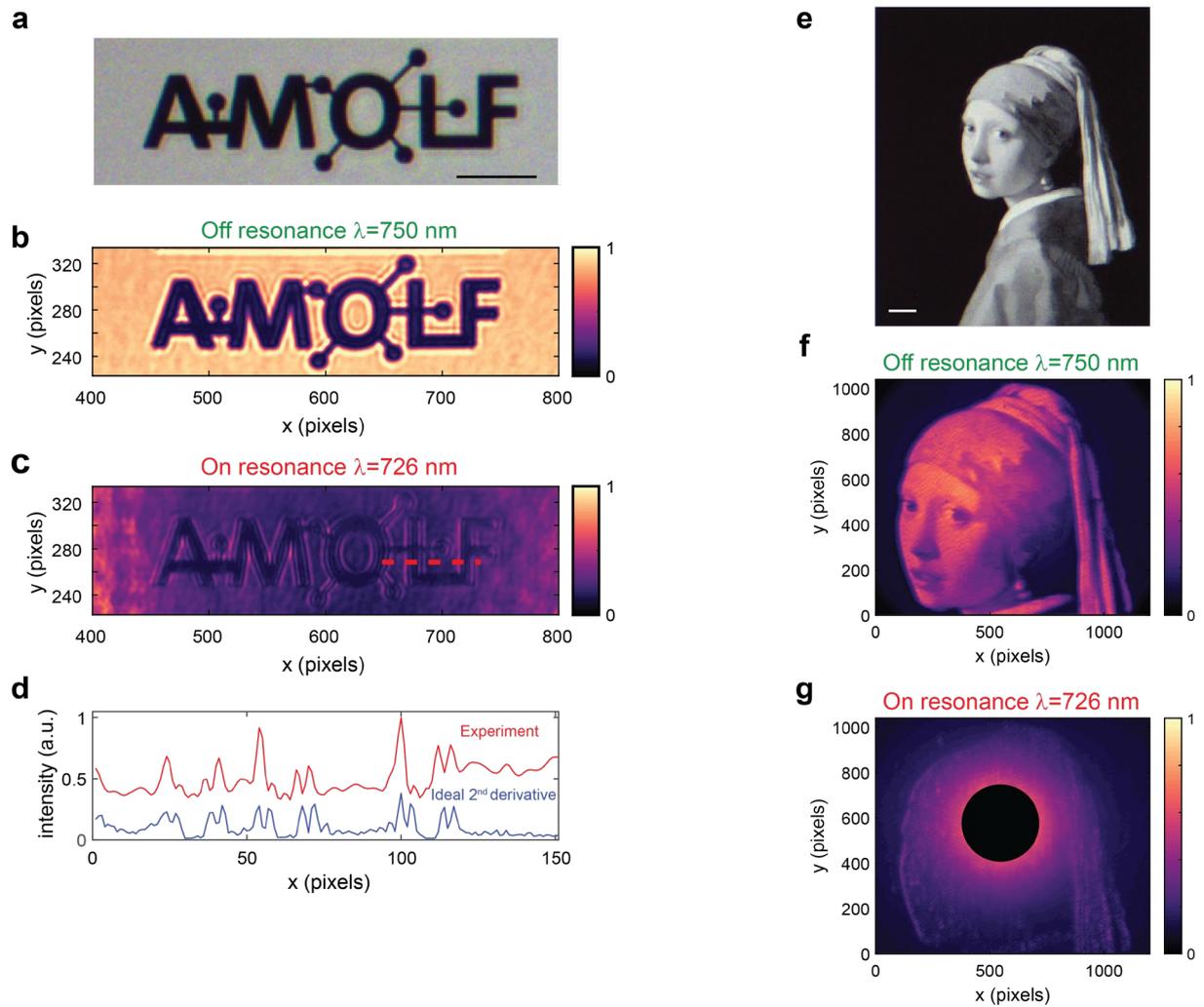

Fig. 5 | **Experimental 2$^{nd}$-order image differentiation**. **a** Optical microscopy image of the input object; the scale bar is 20 μm. **b-c** Optical microscopy image of the metasurface output for resonant ($\lambda$ = 726 nm) and off-resonant ($\lambda$ = 750 nm) illumination. **d** Cross-cut through b (red line) averaged over 8 pixels along y, compared to ideal differentiation performed on the off-resonant image (blue line). **e** Optical microscopy image of the "Meisje met de parel" (J. Vermeer, circa 1665). The image is composed of micron-sized dots of Cr on glass. **f-g** Metasurface output for resonant and off-resonant excitation. The black spot in **g** covers an artefact due to spurious reflection at the interface between air and the sapphire substrate.

To study the edge profile in a quantitative manner, Figure 5d (red curve) shows a line profile taken along the horizontal direction in the processed image (red dashed line in Figure 5c). These data are compared to the calculated output profile assuming an ideal parabolic transfer function (blue curve in Figure 5d). Overall the experimental and ideal response show very similar trends: the double-peaked structure expected for 2$^{nd}$ order differentiation is clearly resolved in all of the six edges shown in Figure 5d.

Finally, we demonstrate the use of a Si metasurface for the processing of a more complex image. We replicated the *Meisje met de parel* painting by J. Vermeer using a suitably designed array of Cr disks on glass (see Figure 5e). An off-resonant transmission image through the metasurface is shown in Figure 5f; the fine features and the contrast in the original object are clearly reproduced in the image. In contrast, the image processed at the resonant wavelength ($\lambda$ = 726 nm) clearly shows the vertical edges along the face contour. The contours are fading away as they become gradually aligned with the x-axis, as expected. This clearly demonstrates that the metasurface image processing concept can be applied to more complex images containing gradients in transmission.

To conclude, we have shown how dielectric metasurfaces sustaining Fano resonances with suitably engineered dispersion can be designed to impart transfer functions in momentum space that correspond to $1^{st}$- and $2^{nd}$-order spatial differentiation. We showed that the ideal amplitude and phase transfer functions can be approximated over a relatively wide range of input angles spanning a numerical aperture up to 0.35 and that transmission over 0.8 can be achieved for large angles. The deviations from the ideal transfer functions, which are intrinsic to the design, are small enough to still achieve derivative operations close to $1^{st}$- and $2^{nd}$-derivative. We experimentally demonstrated the metasurface optical processing using a suitably designed sub-wavelength array of Si nanobeams, showing clear edge detection as a result of the $2^{nd}$-order spatial differentiation. Our 1D design can be readily expanded to 2D operations (e.g. Laplacian) using a 2D array of dielectric nanoparticles. Our results can lead to a wide range of applications and can readily be implemented by directly placing the metasurface onto a standard CCD or CMOS detector chip, opening new opportunities in hybrid optical and electronic computing that operates at low cost, low power, and small dimensions.

## Methods

**Fabrication**

Metasurfaces

The structures were fabricated by Electron Beam Lithography (EBL) as follows:

1) c-Si on $Al_2O_3$ substrates were acquired from MTI corp.

2) The substrate was cleaned in base piranha and the c-Si was etched to the final metasurface thickness via Reactive Ion Etching (RIE) using a two-step process employing $Cl_2$, HBr and $O_2$.

3) The substrate was cleaned again in base piranha and a 200 nm-thick layer of CSAR 62 (AR-P 6200, 9% in anisole) positive-tone resist was spin-coated and baked for 2 minutes at 150˚C.

4) Lines were fabricated in the CSAR layer by exposure using a Raith Voyager lithography system (50 kV, dose 145–150 µC/cm$^2$) and development in Pentyl-acetate (60 s) and o-Xylene (10 s).

5) The pattern was then transferred into the c-Si by a two-step RIE process employing $Cl_2$, HBr and $O_2$.

6) The sample was finally cleaned in anisole at 65˚C followed by an acid piranha cleaning.

Images

1) Glass slides (24×24 mm) were cleaned in base piranha.

2) A bilayer of MMA (MMA(8.5)MAA EL9, 150 nm) and PMMA (PMMA 950k A8, 95 nm) was spin-coated and baked at 150 ˚C and 180 ˚C for 2 minutes respectively.

3) The images were fabricated in the resist layer by exposure using Raith Voyager lithography system (50 kV, dose 550 µC/cm$^2$) and development in MIBK:IPA (1:3 for 90 s).

4) A 40 nm thick Cr layer was evaporated with an in-house built thermal evaporator.

5) The residual resist was lifted-off in anisole at 65˚C.

## Acknowledgments

This work is part of the research program of the Netherlands Organization for Scientific Research (NWO), supported by the European Research Council and by the AFOSR MURI with Grant No. FA9550-17-1-0002.

## Author contributions

AC designed and fabricated Si metasurface samples, performed numerical simulations and performed optical measurements. AC, HK and DS performed theoretical analyses. AFK, AA and AP supervised the project. All authors contributed to the analysis and writing of the paper.

## Competing interests

The authors declare no competing interests.

## ORCID

Albert Polman 0000-0002-0685-3886
Femius Koenderink  0000-0003-1617-5748
Andrea Alu   0000-0002-4297-5274
Dimitrios Sounas 0000-0001-5968-7490
Andrea Cordaro 0000-0003-3000-7943